\documentclass[
preprint,
showpacs,preprintnumbers,
 amsmath,amssymb,
 aps,
]{revtex4-1}

\usepackage{graphicx}
\usepackage{dcolumn}
\usepackage{bm}

\begin{document}


\title{Rotational motion of a droplet induced by interfacial tension}

\author{Ken H. Nagai}
\email[]{nagai@daisy.phys.s.u-tokyo.ac.jp}
\affiliation{Department of Physics, Graduate School of Science, The University of Tokyo, Tokyo, 113-0033, Japan}
\author{Fumi Takabatake}
\affiliation{Department of Physics, Kyoto University, Kyoto, 606-8502,
Japan}
\author{Yutaka Sumino}
\affiliation{Department of Physics, Faculty of Education, Aichi
University of Education, Aichi 448-8542, Japan}
\author{Hiroyuki Kitahata}
\affiliation{Department of Physics, Graduate School of Science, Chiba
University, Chiba 263-8522, Japan}
 \affiliation{PRESTO, Japan Science and
Technology Agency, Saitama, 332-0012, Japan}
\author{Masatoshi Ichikawa}
\affiliation{Department of Physics, Kyoto University, Kyoto, 606-8502,
Japan}
\author{Natsuhiko Yoshinaga}
\affiliation{WPI-AIMR, Tohoku University, Sendai 980-8577, Japan}

\date{\today}

\begin{abstract}
Spontaneous rotation of a droplet induced by the Marangoni flow is
 analyzed in a two-dimensional system. The droplet with the small
 particle which supplies a surfactant at the interface 
 is considered. We calculated flow field around the droplet using Stokes
 equation and found that advective nonlinearity breaks symmetry for
 rotation. Theoretical calculation indicates that the droplet
 spontaneously rotates when the radius of the droplet is an appropriate
 size. 
The theoretical results were validated through comparison with the experiments. 
\end{abstract}

\maketitle

\section{Introduction}
Recently, self-propelled motion of
both biological objects and nonbiological objects are extensively
studied to find general
aspects in self-propelled systems~\cite{Ramaswamy2010,*Hanggi2009}. Collective behaviors such
as nonequilibrium order-disorder transition of direction of motion are
candidates for such general characteristics~\cite{Toner2005}.
One of the best experimental systems to
investigate general aspects is self-propelled objects
driven by interfacial
tension~\cite{Nagai2005,*Nagai2007,*Kitahata2011,Toyota2009,*Thutupalli2011}. Actually,
several results on collective behavior of the objects driven by
interfacial tension have been reported recently~\cite{Suematsu2010,*Soh2008}. This is because this kind of
systems usually consists of a small number of elements and allows us to
perform experiments under broad parameter space. Despite of this
advantages, however, mechanical analysis for above droplet motion is
still lacking.

The above-described droplet motion driven by interfacial tension  is observed
even under isotropic conditions. Although a droplet itself has no asymmetry, symmetry of interfacial tension is
broken through the advective nonlinearity. Actually, Toyota~{\it et~al}. reported spontaneous translational
motion of a circular droplet~\cite{Toyota2009,*Thutupalli2011}. Recently,
theoretical analysis for such translational motion has been reported and the nonlinear
effect which breaks the symmetry has been made
clear~\cite{Yabunaka2012,*Yoshinaga2012}. Not only spontaneous translational motion
but also spontaneous
rotation has also been demonstrated by
Takabatake {\it et~al}. experimentally~\cite{Takabatake2011}. In this
system, a droplet rotates even though the system is mirror
symmetric. Although this system was analyzed using Langevin equation in
the previous study, mechanical analysis is lacking and the physical mechanism of
symmetry breaking is not clear. In this article, we analyzed the
spontaneous rotation of a droplet using an advection-diffusion equation
coupled with
Stokes equation. We found that the droplet can rotate when the droplet
has an appropriate radius. To validate our model, theoretical results were compared with the
experimental ones.

\section{model equation}
\begin{figure}
\includegraphics{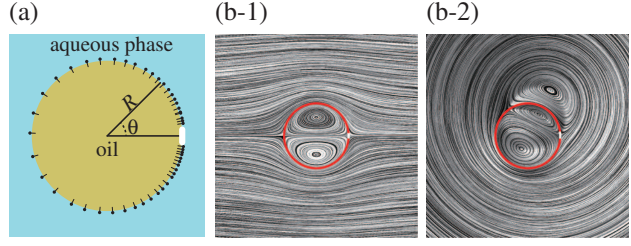}
\caption{(Color Online) (a) Schematic diagram of the model in this
 article. (b) The flow field when the interface concentration field is
 $c(\theta)$. The images are stream line
 around the droplet in the frame where the particle and the center of
 mass of the droplet are fixed. The red line is the oil-water interface
 and the white
 line is the particle. (b-1) is the flow field when $c(\theta)=1+0.1 \cos\theta$ (symmetric
 concentration field) and (b-2) is the flow field when
 $c(\theta)=1+0.2\cos\theta +0.1 \sin\theta$ (asymmetric concentration field).}
\label{schematic}
\end{figure}

We consider a circular undeformable oil droplet with a radius, $R$, in
an aqueous phase
in a two-dimensional space, as shown in Fig.~\ref{schematic}(a). A small solid
particle is attached at the interface and a surfactant that reduces
interfacial tension is supplied to
both the inside  and the interface of the droplet from
the particle. In the droplet, the surfactant is decomposed. As a result
of these process, there appears inhomogeneity of interfacial tention, and the droplet is driven
by the Marangoni flow induced by the interfacial tension gradient. In a laboratory frame, the
particle can freely move along the interface. Without loss of
generality, we may also consider a frame in which the
particle and the center of mass of the droplet are fixed. Polar
coordinates are introduced with the center of
mass of the droplet regarded as the origin. The position of the particle is fixed at
$r=R$ and $\theta = 0$. 
With the assumption of low Reynolds number, the velocity field is
calculated using Stokes equation as follows,
\begin{equation}
 -\boldsymbol{\nabla} p + \eta\boldsymbol{ \nabla}^{2}  \bm{v}=\bm{0}, \label{stokes}
\end{equation}
where $\bm{v}$ is the velocity, $p$ is the
pressure field and $\eta$ is the viscosity coefficient. In our model, $\bm{v}$
satisfies the incompressive condition, $\boldsymbol{\nabla} \cdot \bm{v}
= 0$. Due to the interfacial gradient, there is the following stress jump at the interface~\cite{Young1959}:
\begin{equation}
\left. \sigma^{{\rm (i)}}_{r\theta} \right|_{r=R} = \left. \sigma^{{\rm
 (o)}}_{r\theta}\right|_{r=R} + \frac{1}{R}\frac{\partial
\gamma}{\partial \theta}, \label{stressjump}
\end{equation}
where $\bm{\sigma}$ is the stress tensor, the superscripts ``(i)'' and
``(o)'' correspond to the inside and
outside of the droplet, and $\gamma(\theta)$ is
the interfacial tension between two fluids. Here, $\gamma$ linearly depends on the
interface concentration of the surfactant, $c(\theta)= c^{0}+ \sum_{m=1}^{\infty}
 \left(c_{m}^{\text{c}}\cos m \theta +
 c_{m}^{\text{s}} \sin m \theta\right)$, as $\gamma(\theta) =
 \gamma_{0}-k c(\theta)$.

Assuming that the surfactant supplied from the particle is  distributed only at the interface due
to fast decomposition in the bulk, the time evolution of $c(\theta)$ can
be described as
\begin{align}
\frac{\partial c}{\partial t} + \frac{u}{R}
\frac{\partial c}{\partial \theta} = \frac{D}{R^{2}} \frac{\partial^{2}
 c}{\partial \theta^{2}} - \alpha c + \pi \alpha \beta \delta(\theta),\label{eq-1}
\end{align} 
where $D$ is the diffusion constant at the interface, $\alpha$ is the
decomposition rate, and $\pi\alpha\beta$ is the rate of supply. In Eq.~\eqref{eq-1}, the stationary flow field satisfying
Eq.~\eqref{stokes} is used as $u$, which is determined by
$c(\theta)$.

\section{Solution of the two dimensional Stokes equation}
In this section, $\bm{v}$ satisfying Eq.~\eqref{stokes} is calculated under
given concentration field at the interface, $c(\theta)$.
In the polar coordinate, $\bm{v}$ is described as
\begin{equation}
 \bm{v}=v_{r}(r, \ \theta)\bm{e}_{r} + v_{\theta}(r, \ \theta)\bm{e}_{\theta},
\end{equation}
where $\bm{e}_{r}$ and $\bm{e}_{\theta}$ are the unit vectors in the
radial and angular directions, respectively.
Due to incompressibility, 
\begin{equation}
 \boldsymbol{\nabla} \cdot  \bm{v} = \frac{1}{r} \frac{\partial}{\partial
  r}\left( r v_{r} \right) +
\frac{1}{r}  \frac{\partial v_{\theta}}{\partial \theta}=0;
\end{equation}
therefore, using the stream function,
$\Psi$, the solution of Eq.~\eqref{stokes} can be described as
\begin{align}
 v_{r} =& \frac{1}{r} \frac{\partial \Psi}{\partial \theta},\label{222034_6Nov11}\\
 v_{\theta} =& - \frac{\partial \Psi}{\partial r}.\label{222050_6Nov11}
\end{align}
Taking
rotation of both sides of
Eq.~\eqref{stokes} and substituting Eqs.~\eqref{222034_6Nov11}
and \eqref{222050_6Nov11}, we can obtain
\begin{equation}
 \boldsymbol{\nabla}^{2}\boldsymbol{\nabla}^{2} \Psi = 0,\label{012331_30Oct11}
\end{equation}
where
\begin{equation}
 \boldsymbol{\nabla}^{2}=\frac{1}{r}\frac{\partial}{\partial r} \left(r
							    \frac{\partial}{\partial
							    r}\right)+\frac{1}{r^{2}}\frac{\partial^{2}}{\partial
 \theta^{2}}.
\end{equation}
The solutions inside the droplet, $\Psi^{\rm (i)}$, and the outside the
droplet, $\Psi^{\rm (o)}$, are
\begin{align}
 \Psi^{\rm (i)} (r, \ \theta) =& B_{0}^{\rm (i)} \left(\frac{r}{R}\right)^{2}   \notag\\
 & +\left(  A_{1}^{\rm c}
 \left(\frac{r}{R}\right)^{3} + C_{1}^{\rm c(i)} \frac{r}{R}  \right)
 \cos \theta \notag\\ 
&+  \left(  A_{1}^{\rm s}
 \left(\frac{r}{R}\right)^{3} + C_{1}^{\rm s(i))} \frac{r}{R} \right)  \sin  \theta \notag\\
 & + \sum_{m=2}^{\infty} \biggl\{ \left( A_{m}^{\rm
 c}\left( \frac{r}{R}  \right)^{m+2}  + C_{m}^{\rm c}\left(
 \frac{r}{R}\right)^{m}  \right)\cos m \theta\notag\\
&+ \left( A_{m}^{\rm s} \left( \frac{r}{R}
 \right)^{m+2} + C_{m}^{\rm s}\left( \frac{r}{R}\right)^{m} 
 \right) \sin m \theta   \biggr\},\\
 \Psi^{\rm (o)} (r, \ \theta) =&  B_{0}^{\rm (o)} \left(\frac{r}{R}\right)^{2} + C_{0}\ln \frac{r}{R}  \notag\\
&+  \left(  B_{1}^{\rm c} \frac{r}{R} \ln \frac{r}{R} + C_{1}^{\rm c(o)} \frac{r}{R} + D_{1}^{\rm c} \frac{R}{r}      \right) \cos \theta\notag\\
&+  \left(  B_{1}^{\rm s} \frac{r}{R} \ln \frac{r}{R} + C_{1}^{\rm s(o)} \frac{r}{R} + D_{1}^{\rm
 s} \frac{R}{r}    \right)  \sin  \theta \notag\\
 & + \sum_{m=2}^{\infty} \biggl\{ \left(  B_{m}^{\rm c}
 \left( \frac{R}{r}  \right)^{m-2}  + D_{m}^{\rm c} \left(\frac{R}{r}\right)^{m}  \right)\cos m \theta\notag\\
& + \left( B_{m}^{\rm s} 
 \left( \frac{R}{r}  \right)^{m-2} + D_{m}^{\rm s}\left(\frac{R}{r}\right)^{m}
 \right) \sin m \theta   \biggr\}.
\end{align}

The boundary conditions of flow field at the interface are $\left.v_{r}^{\rm
(o)}\right|_{r=R}=\left.v_{r}^{(i)}\right|_{r=R}=0$ and
$\left.v_{\theta}^{\rm
(o)}\right|_{r=R}=\left.v_{\theta}^{(i)}\right|_{r=R}$ since the droplet
shape is fixed.
Since there is no external force, a torque free condition,
\begin{equation}
 \bm{T} =\int {\rm d}l \left(\bm{n}\cdot \left. \bm{\sigma}^{\rm
 (o)}\right|_{r=R} \right)\times \bm{r}=\bm{0},
\end{equation}
and a force free condition,
\begin{equation}
 \bm{f} = \int {\rm d}l \bm{n}\cdot \left.\bm{\sigma}^{\rm
 (o)}\right|_{r=R}=\bm{0},
\end{equation}
have to be satisfied,
where $l$ is the distance along the interface.
Using the conditions described above and Eq.~\eqref{stressjump}, the
coefficients are calculated as
\begin{align}
 B_{0}^{\rm (i)}=&B_{0}^{\rm (o)},\notag\\
C_{0}=&0,\notag\\
B_{1}^{\rm c}=&B_{1}^{\rm s}=0,\notag\\
 A_{1}^{\rm c}=&-C_{1}^{\rm c(i)}=C_{1}^{\rm c(o)}=-D_{1}^{\rm
 c}=\frac{\kappa c_{m}^{\rm s}R}{2},\notag\\
 A_{1}^{\rm s}=&-C_{1}^{\rm s(i)}=C_{1}^{\rm s(o)}=-D_{1}^{\rm
 s}=-\frac{\kappa c_{m}^{\rm c}R}{2},\notag\\
 A_{m}^{\rm c}=&B_{m}^{\rm c}=-C_{m}^{\rm c}=-D_{m}^{\rm
 c}=\frac{\kappa c_{m}^{\rm s} R}{2}\notag\\
&(m\neq 0,1),\notag\\
 A_{m}^{\rm s}=&B_{m}^{\rm s}=-C_{m}^{\rm s}=-D_{m}^{\rm
 s}=-\frac{\kappa c_{m}^{\rm c}R}{2}\notag\\
&(m\neq 0,1),\label{coefficient}
\end{align}
where $\kappa=k/\left\{2 \left(\eta^{(\text{o})} +
\eta^{(\text{i})}\right)\right\}$.
From Eqs.~\eqref{222034_6Nov11}, \eqref{222050_6Nov11} and \eqref{coefficient}, flow profiles
are obtained as
\begin{align}
 v_{r}^{\rm (o)}=&-\sum_{m=1}^{\infty}\frac{\kappa m \left[\left(\frac{R}{r}\right)^{m-1} -
 \left(\frac{R}{r}\right)^{m+1}\right]}{2} \notag\\
&\qquad\qquad  \left( c_{m}^{\rm c}\cos
 m\theta + c_{m}^{\rm s} \sin m \theta   \right),\notag\\
 v_{r}^{\rm (i)}=&- \sum_{m=1}^{\infty} \frac{\kappa m \left[ \left(\frac{r}{R} \right)^{m+1}-
 \left(\frac{r}{R}\right)^{m-1}  \right] }{2} \notag\\ 
&\qquad\qquad \left( c_{m}^{\rm c} \cos m \theta + c_{m}^{\rm s} \sin m \theta \right),\notag\\
 v_{\theta}^{\rm (o)}=&- 2 \frac{B_{0}^{\rm (i)}}{R^{2}} r \notag\\
& - \sum_{m=1}^{\infty} \frac{\kappa\left[ (m-2) \left( \frac{R}{r}
 \right)^{m-1}  -m \left( \frac{R}{r}  \right)^{m+1} \right] }{2}\notag\\
&\qquad\qquad \left(
 c_{m}^{\rm c} \sin m\theta - c_{m}^{\rm s} \cos m \theta  \right),\notag\\
 v_{\theta}^{\rm (i)}=&-2 \frac{B_{0}^{\rm (i)}}{R^{2}}r\notag\\
& - \sum_{m=1}^{\infty} \frac{\kappa\left[ -(m+2) \left( \frac{r}{R}
 \right)^{m+1} + m\left( \frac{r}{R} \right)^{m-1}  \right]}{2}\notag\\
&\qquad\qquad \left(
 c_{m}^{\rm c} \sin m\theta - c_{m}^{\rm s} \cos m \theta  \right),
\end{align}
and the tangential flow at the interface $(r=R)$, $u(\theta)$, is
obtained as
\begin{equation}
u(\theta) = -2\frac{B_{0}^{\rm (i)}}{R^{2}} R + \kappa \sum_{ m = 1 }^{ \infty } \left( c_{m}^{\text{c}} \sin m\theta - c_{m}^{\text{s}} \cos m\theta 
\right). \label{flow}
\end{equation}

As the particle is a solid, the flow velocity on the particle is
$\bm{0}$ (no-slip condition). Assuming that the particle size is
small enough, this condition can be described as $u(0) = 0$. Using this
condition, $B_{0}^{\rm (i)}$ is determined as
\begin{equation}
B_{0}^{\rm (i)} =- \frac{\kappa R^{2}}{2} \sum_{ m = 1 }^{ \infty } \frac{ c_{ m }^{\text{s}}}{ R}.\label{omega}
\end{equation}
At infinity, $v_{r}^{\rm (o)}=-\kappa \left(c_{1}^{\rm c}\cos \theta +
 c_{1}^{\rm s}\sin \theta\right)/ 2$ and $v_{\theta}^{\rm (o)}=-2\frac{B_{0}^{\rm
 (i)}}{R^{2}}r - \kappa
\left(-c_{1}^{\rm c}\sin\theta + c_{1}^{\rm
 s}\cos\theta\right)/2$, which
 means the droplet translational velocity ($\bm{V}$) and the angular velocity in the
 laboratory frame ($\Omega$) are
\begin{align}
\bm{V} =& \kappa\frac{c_{1}^{\text{c}}}{2} \bm{e}_{x}+  \kappa\frac{c_{1}^{\text{s}}}{2}  \bm{e}_{y},\label{162939_26Oct11}\\
\Omega =&2 \frac{B_{0}^{\rm (i)}}{R^{2}}=-\kappa\sum_{m=1}^{\infty}\frac{c^{\text{s}}_{m}}{ R
}.\label{162949_26Oct11}
\end{align}
$\bm{e}_{x}$ and $\bm{e}_{y}$ in Eq.~\eqref{162939_26Oct11} are the unit vectors in the
direction of $x$ axis ($\theta=0$) and $y$ axis ($\theta=\pi/2$),
respectively. The calculated flow fields are illustrated in the case of
translational motion and in the case of
rotational motion using Line Integral Convolution~\cite{Cabral1993} in Fig.~\ref{schematic}~(b).
It is noted that the rotational speed of a droplet is proportional to the
sum of the Fourier coefficients corresponding to an antisymmetric
concentration field about the anterior-posterior axis. 

\section{Analysis of the model}
Using Eq.~\eqref{eq-1}, we calculate the velocity and the angular
velocity of the droplet in this section.
Since higher Fourier modes of $c$ decay quickly due to
diffusion, we neglect $c_{n}^{\text{c}}$,
and $c_{n}^{\text{s}}$ for $|n| \geq
2$. Letting $c_{0}^{\text{c}}$, $c_{1}^{\text{c}}$, and
$c_{1}^{\text{s}}$ be replaced by $X$, $Y$ and $Z$, Eq.~\eqref{eq-1} is described as 
\begin{align}
 \frac{\text{d} X}{\text{d}t}=& - \alpha (X - X_{0} ) + \frac{\kappa}{ R
 } \left\{ Y^{2} -\left( Y_{0}\right)^{2} + Z^{2} \right\}, \label{112348_14May11}\\
 \frac{\text{d} Y}{\text{d} t} =& \lambda_{Y} (Y - Y_{0}) - \frac{\kappa}{R} Z^{2},\label{233955_30Mar11}\\
 \frac{\text{d}Z}{\text{d} t} =& \lambda_{Z} Z + \frac{\kappa}{ R} \left(Y-Y_{0}\right) Z.\label{234003_30Mar11}
\end{align}
Here, $X_{0}=\kappa
 \beta^{2}\rho^{2}/R \left(\rho^{2}+1\right)  + \beta/2,$ and
 $Y_{0}=\beta\rho^{2}/\left(\rho^{2}+1\right)$ are the values of $X$ and
 $Y$ at the fixed 
point at which $Z=0$. Here, $\rho=R/l_{\rm d}$ is the radius normalized by the
 characteristic length 
 of diffusion, $l_{\rm d}=\sqrt{D/\alpha}$.
$\lambda_{Y}$ and $\lambda_{Z}$ are the eigenvalues for the linearized
 equations for $Y$ and $Z$, 
\begin{align}
\lambda_{Y}=&-\alpha\left(1+\frac{1}{\rho^{2}}\right),\\
\lambda_{Z}=&\frac{\kappa}{R} Y_{0}- \alpha
 \left(1+\frac{1}{\rho^{2}}\right) = \frac{\alpha \rho^{2}}{\rho^{2}+1}
 \left\{\frac{ L} {\rho} - \left(1+\frac{1}{\rho^{2}}\right)^{2}
 \right\},\label{195007_20Oct11}
\end{align}
respectively, where $L=l_{\rm a}/l_{\rm d}$ is the ratio between another characteristic
length, $l_{\rm a}=\kappa \beta/\alpha$, and $l_{\rm d}$. $l_{\rm a}$ is the length of
advective transportation during $1/\alpha$ since $u\sim \kappa c_{1}^{\rm
c} \sim \kappa\beta \rho^{2} / \left( \rho^{2} + 1 \right) \sim \kappa
\beta$. 
If $\lambda_{Z}$ is positive, the fixed point, $(X_{0},Y_{0},0)$, is
unstable, which means a finite angular velocity of a droplet obtained from
Eq.~\eqref{162949_26Oct11}.

\begin{figure}
\includegraphics{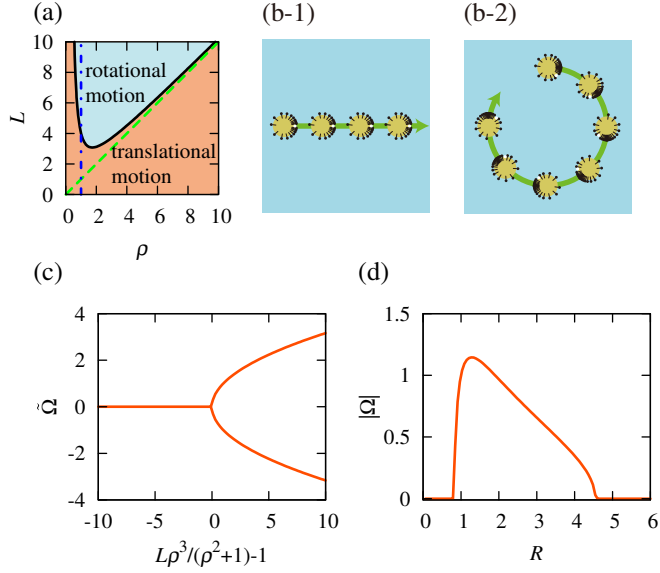}
\caption{(Color Online) Bifurcation of the motion of a droplet.  (a) Phase diagram of the motion. One-point
 dashed line (blue) and dashed line
 (green) represent $R=l_{\rm d}$ and
 $R=l_{\rm a}$, respectively. (b) Schematic diagrams of the interface
 concentration field and the motion of a droplet in a laboratory
 frame. When the droplet moves straight (b-1) and rotates (b-2), the
 concentration field is mirror symmetric and asymmetric about the
 anterior-posterior axis, respectively. (c) Dependence
 of $\tilde{\Omega}$ on $L\rho^{3}/\left(\rho^{2} +
 1\right)^{2}-1$. (d) Dependence of $|\Omega|$ on
 $R$. $\alpha=1$, $l_{\rm d}=1$, and $l_{\rm a}=5$. With these
 values, $R_{\rm
 l}^{\rm c}=0.83$, $R_{\rm u}^{\rm c}=4.55$ and $|\Omega|$ reaches a
 maximum value at $R=1.28$.}
\label{200959_6Apr11}
\end{figure}

The phase diagram of the
 motion can be illustrated as shown in Fig.~\ref{200959_6Apr11} (a) using two
 kinds of motion: translational motion and rotational motion. To realize rotation, it
is necessary that the advection term in Eq.~\eqref{eq-1} is dominant;
that is, $L \gtrapprox 1$ is needed. In fact, $\lambda_{Z}$ can be positive only when
$L>16\sqrt{3}/9\approx 3.08$. When $R\lessapprox l_{\rm d}$ ($\rho \lessapprox 1$), the
concentration field is almost uniform due to diffusion,
so that flow is too weak to generate rotation. On the other hand, when
$R\gtrapprox l_{\rm a}$ ($\rho\gtrapprox L$ ), the
uniform decomposition term in \eqref{eq-1}, $-\alpha c$, becomes
dominant and
the concentration field is almost uniform. Therefore, only when $L$ is large
enough and $R$
 is between $R_{\rm l}^{\rm c}\approx l_{\rm d}$ and $R_{\rm u}^{\rm
 c}\approx l_{\rm a}$, can the droplet 
 spontaneously rotate due to the coupling between the mirror symmetric
 concentration field ($Y_{0}$) and mirror
 antisymmetric concentration field ($Z$) through the rotational flow
 field. Since there is no other bifurcation, there are only a pair of
 stable fixed 
 points with non-zero
 $Z$, which are calculated as $X=3 \alpha l_{\rm a}/2 \kappa$,
 $Y= -R \lambda_{Y} /\kappa$ and $Z=\pm R
 \sqrt{-\lambda_{Y}\lambda_{Z}}/\kappa$, when $R_{\text{l}}^{\text{c}}<
 R <R_{\text{u}}^{\text{c}}$. Using
Eq.~\eqref{162939_26Oct11} and Eq.~\eqref{162949_26Oct11},  $\bm{V}$ and
$\Omega$ in the steady state are calculated as
\begin{align}
\bm{V} &= 
\begin{cases}
 \left(-\lambda_{Y}l_{\rm a}, \ 0\right)&( R<
 R_{\text{l}}^{\rm c}, R> R_{\text{u}}^{\text{c}} )\\
\left(-\frac{\lambda_{Y} R }{2}, \ \pm
 \frac{R}{2}
 \sqrt{-\lambda_{Y}\lambda_{Z}} \right) & (
R_{\text{l}}^{\text{c}} \leq R \leq R_{\text{u}}^{\rm c})
\end{cases},
\\
\Omega &= 
\begin{cases}                                                                 
 0 & ( R< R_{\text{l}}^{\rm c},  R> R_{\text{u}}^{\text{c}} )\\
\mp    \sqrt{-\lambda_{Y}\lambda_{Z}}\\
\quad =\mp \alpha
 \sqrt{\frac{l_{\rm a}}{R} - \left\{1 + \left(\frac{l_{\rm d}}{R}\right)^{2}  \right\}^{2}} &(
R_{\text{l}}^{\text{c}} \leq R \leq R_{\text{u}}^{\rm c})
\end{cases}.\label{Omega}
\end{align}
The interface concentration and the motion
of the droplet in a laboratory frame are schematically illustrated in
Fig.~\ref{200959_6Apr11}~(b).
When a droplet exhibits translational motion ($R<R_{\rm l}^{\rm c},R>R_{\rm u}^{\rm c}$), the interface concentration is symmetric
about the anterior-posterior axis and the anterior-posterior axis always
corresponds to the direction of the motion. On the other hand, when a droplet
rotates ($R_{\rm l}^{\rm c}<R<R_{\rm u}^{\rm c}$), the interface
concentration is asymmetric about the anterior-posterior axis and the
particle is always the inside of the trajectory. Considering the normalized $\Omega$,
$\tilde{\Omega}=\Omega /\lambda_{Y}$,
$\tilde{\Omega}$ is the square root of the distance from the critical point, $|L
\rho^{3}/ \left( \rho^{2} + 1 \right)^{2} -1|$, which corresponds to the
stable solution of the normal form of pitch-fork
bifurcation~\cite{Guckenheimer1990}, as shown in
Fig.~\ref{200959_6Apr11}~(c).  Figure~\ref{200959_6Apr11}~(d) shows the
dependence of $|\Omega|$ on $R$ when  $l_{\rm d}$, $l_{\rm a}$ and $\alpha$ are fixed. $|\Omega|$ has a maximum 
value at a radius between $R_{\rm l}^{\rm c}$ and $R_{\rm u}^{\rm c}$.

\section{Comparison with the Experiment}
\begin{figure}
 \includegraphics{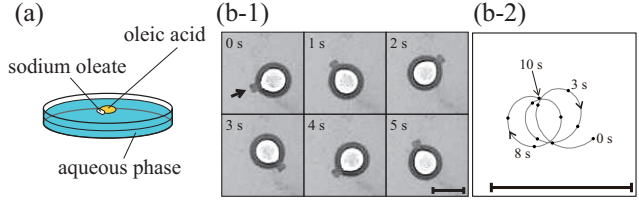}
\caption{(Color Online) Outline of the experimental system. (a) Schematic diagram. A petri dish with a radius of 7.5~cm was filled with
 100~ml water. A droplet of oleic acid together with a 3~mm particle of
 sodium oleate was placed on the water surface. The droplet moved spontaneously
 driven by the Marangoni effect~\cite{Takabatake2011}. (b) Trajectory of
 a droplet with a 
 radius of 5.9~mm (300~$\mu$l). (b-1) Snapshots of a droplet per
 1~s. The arrow indicates the particle fixed on the
 interface.  (b-2) Trajectory of the center of
 mass of the droplet. Both scale bars represent 10~mm. }
\label{233253_14May11}
\end{figure}

Using above theoretical results, we analyzed the corresponding
experiment reported in~\cite{Takabatake2011} with additional data. The schematic diagram of
the experimental setup is
illustrated in Fig.~\ref{233253_14May11}~(a). One hundred milliliters of water, which was
purified with a MilliQ filtering system (Millipore), was placed to a
petri dish.  A droplet of oleic acid
(Wako Pure Chemical Industries; 159-00246) and a solid sodium oleate (soap) were
floated on the aqueous phase in the petri dish.  A solid column of sodium
oleate (3~mm in length) was chosen from a commercial sample
(Nacalai Tesque; 257-02). The movement of
the solid/liquid composite was captured by a digital video camera at 30
frames per second at room temperature, and then analyzed using Image J (http://rsbweb.nih.gov/ij/docs/index.html).

\begin{figure}
 \includegraphics{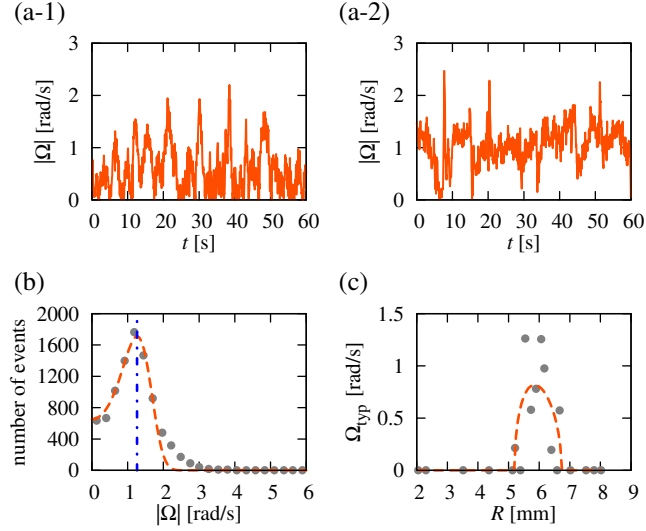}
\caption{(Color Online) Analysis of the experimental results. (a) Time
 series of the absolute value of angular velocity of a droplet with a
 radius of 2.3~mm (50~$\mu$l; a-1) and 
  5.6~mm (260~$\mu$l; a-2). (b) Distribution of $\left|\Omega\right|$ obtained from
 experiments using a droplet with a radius of 5.6~mm (closed circles). Dashed line (orange) is
 the fitted curve. The parameters were estimated as
 $a=3.19$,
 $b=0.78$ and $c=6.48$. One-point dashed line (blue) represents $\Omega_{\rm
 typ}=1.3$~rad/s.  (c) Droplet size dependence of $\Omega_{\rm typ}$. Gray circles represent $\Omega_{\rm typ}$. Dashed line (orange) is the fitted
 curve. Estimated parameters were $\alpha=5.4$~s$^{-1}$, $l_{\rm d}=3.4$~mm
 and $l_{\rm a}=10.6$~mm.}
\label{183302_28Sep11}
\end{figure}

A time series of snapshots of a droplet, and the corresponding trajectory of the center of mass of the droplet are
shown in Fig.~\ref{233253_14May11}~(b). Due to the fluctuation of
rotational speed, the trajectory does not exhibit a closed circle. From the analysis of observed
trajectories, the angular velocity, $\Omega$, was
measured. Time
courses of $|\Omega|$ are shown in Fig.~\ref{183302_28Sep11}~(a). Five
experiments were performed for each volume to yield the distribution of $|\Omega|$, which is shown in Fig.~\ref{183302_28Sep11}~(b).
To determine the maximum value of the distribution, the distribution was
fitted with the function, $\exp\left\{ -b^{4}\Omega^{2}\left(\Omega^{2} -a \right)+c
\right\}$,  using the weighted nonlinear least-squares Levenberg-Marquardt
algorithm, where the fitting parameters were $a$, $b$ and $c$~\cite{Levenberg1944}. The
number of events was taken as the weight for fitting. Using these
parameters, the peak of the distribution, $\Omega_{\rm typ}$, is calculated as 0 when $a<0$, and
$\sqrt{a/2}$ when $a\geq 0$. We used $\Omega_{\rm typ}$ as the typical value
of $\Omega$. To see the dependence of $\Omega_{\rm typ}$ on the radius,
$\Omega_{\rm typ}$ is plotted against the droplet radius in
Fig.~\ref{183302_28Sep11}~(c). Here, the radius was
estimated from the image of the droplet. As shown in Fig.~\ref{183302_28Sep11} (c), the radius with the highest $\Omega_{\rm typ}$ and the critical
radii, $R_{\rm l}^{\rm c}$ and $R_{\rm u}^{\rm c}$, were all observed as predicted by our
theoretical considerations. By fitting  Eq.~\eqref{Omega} to the obtained points, $(R,
\ \Omega_{\text{typ}})$, we estimated $l_{\rm d}$,
$l_{\rm a}$ and $\alpha$. 
The estimated parameters of the fitted curve, which is shown in Fig.~\ref{183302_28Sep11}~(c), were $\alpha=5.4$~s$^{-1}$,
 $l_{\rm a}=10.6$~ mm, and $l_{\rm d} = 3.4$~mm. It is noted that the time scale
 (the period of rotation) and the space scale (the size of the droplet and the
 particle) in the experiment are
 the same order as $1/\alpha$, and $l_{\rm d}$ and $l_{\rm a}$,
 respectively. So far,
quantitative comparison is not possible within our model
since $\alpha$ and $\beta$ is difficult to measure experimentally.

\section{Summary}
In summary, we manifested the physical mechanism of the spontaneous rotation of
a mirror symmetric system consisting of a droplet and a particle
attached at the interface. Using the flow field calculated with Stokes equation, we 
analytically found that when the Marangoni flow is strong enough, there
are two critical radii of the droplet 
for rotation and the peak of angular velocity at a certain radius between these two
critical radii. For verification, we compare our model with the
experiment and these theoretical predicted results  were ascertained in
the experiment. Although we analyze only the spontaneous rotation driven
by the Marangoni effect in this article, the advective transport
of objects is generally observed in self-propelled phenomena such as droplets driven by
any propulsive force and amoebae cells. Therefore, the advective
nonlinearity similar to that in this article is expected to cause the symmetry breaking for rotation in
various kinds of self-propelled phenomena.

\section{Acknowledgement}
The authors thank T. Ohta, K. Yoshikawa, M. Sano, and H. Kori 
for their helpful discussion. K. H. N. is supported by a JSPS fellowship
for young scientists (No.23-1819). This study was supported by  Grants-in-Aid
for Young Scientists (B) (No.21740282) to H. K. and (No.23740317) to N. Y.



%

\end{document}